\documentclass[a4paper]{article}
\usepackage{fullpage}
\usepackage[utf8]{inputenc}

\usepackage{array}
\usepackage{verbatim}
\usepackage[]{amsmath}
\usepackage{amssymb}
\usepackage{setspace}
\usepackage[]{graphicx}
\usepackage{subfig}
\doublespace

\newcommand{\ket}[1]{\vert #1 \rangle}
\newcommand{\braket}[2]{\langle #1 \vert #2 \rangle}
\newcommand{\bra}[1]{\langle #1 \vert}
\newcommand{\lrangle}[1]{\langle #1 \rangle}

\DeclareMathOperator{\tr}{tr}

\title{New Supercoherent States}
\author{Mordechai Kornbluth, Fredy Zypman \\ Yeshiva University, Physics Department \\ 500 W 185th Street, New York NY 10033 \\ mkornblu@yu.edu, zypman@yu.edu }

\begin{document}

\maketitle

\begin{abstract}
This study generalizes the supersymmetric coherent states introduced by Aragone and Zypman in Ref. \cite{Aragone}. The Hamiltonian of the supersymmetric quantum harmonic oscillator leads to the definition of the generalized supersymmetric annihilation operators as a 3-parameter family. Their eigenstates are the generalized supercoherent states, which can be calculated explicitly for three relevant regions of $\vec{k}$ space. The uncertainty in position and momentum is discussed, with specific concentration on where the uncertainty is saturated, where it is bounded, and where it is unbounded.
\end{abstract}

\vspace{36pt}

``Of systems possible, if 'tis confest, that wisdom infinite must form the best, where all must fall or not coherent be, and all that rises, rise in due degree.'' --Alexander Pope, \textit{Essay on Man}

\section{Introduction}

Coherent states have attracted attention in recent decades, especially in signals processing and field theory; for a review of research on coherent states, see Ref. \cite{Dodonov}. The ``canonical'' coherent states for the quantum-mechanical harmonic oscillator have three definitions, which are equivalent for the harmonic oscillator potential: (1) They are minimum-uncertainty states (MUS, $\sigma_x \sigma_p = \hbar/2$), (2) They are eigenstates of the annihilation (lowering) operator $\hat{a}$, which lowers a state of definite energy $E_n$ to one of $E_{n-1}$ ($\hat{a} \ket{\alpha} = \alpha \ket{\alpha}$), and (3) they are equal to the displacement operator acting on the ground state ($\ket{\alpha} = \exp(\alpha \hat{a}^+ - \alpha^* \hat{a}) \ket{0}$).

For non-harmonic Hamiltonians, finding MUS is nontrivial. Many authors search for MUS by starting with ``generalized coherent states'' defined with one of the other two definitions (see Ref. \cite{Nieto}). We take this approach, extending definition (2) to investigate coherent states of the supersymmetric harmonic oscillator.

Aragone and Zypman (Ref. \cite{Aragone}) first introduced the supersymmetric coherent states for the harmonic oscillator. Recent description of the motivation for studying supersymmetric quantum mechanics can found in Ref. \cite{Lee}; see there for particular discussion of the application of supersymmetric quantum mechanics to solve partner Hamiltonians (see also Ref. \cite{Fernandez}). This research uses the Hamiltonian of Ref. \cite{Aragone}, for other harmonic-oscillator Hamiltonians, see Ref. \cite{Berube}, \cite{Cherbal}. Similar research has analyzed non-harmonic Hamiltonians; see Ref. \cite{Angelova}. For a different treatment of supercoherent states, see Ref. \cite{Fatyga}.

The simplest supersymmetric annihilation operator (SAO) is the trivial extension of the bosonic annihilation operator into superspace,
\begin{equation}
  \hat{A}_t =
  \begin{pmatrix}
    a & 0 \\ 0 & a
  \end{pmatrix}
  \label{eq:Superannihilation_Trivial}
\end{equation}

However, $\hat{A}_t$ does not mix the bosonic and fermionic components; to wit, for any superstate $
\begin{pmatrix}
  \ket{\Psi_b} & \ket{\Psi_f}
\end{pmatrix}^T
$, diagonalized operators do not mix the components $\ket{\Psi_b}$ and $\ket{\Psi_f}$. 

Ref. \cite{Aragone} identified one nontrivial SAO $\hat{A}_a$, which mixes bosonic and fermionic components. 
\begin{equation}
  \hat{A}_a = 
  \begin{pmatrix}
    a & 1 \\ 0 & a
  \end{pmatrix}
  \label{eq:Superannihilation_Aragone}
\end{equation}

They defined its eigenstates as ``supercoherent states,'' following the above definition (2) of the coherent state. This work shows that $\hat{A}_a$ is only one example of the family of SAOs; we then analyze the generalized SAO and its eigenstates (supercoherent states). 

The following notational conventions are taken in this paper: (1) As in Ref. \cite{Aragone}, supercoherent states are defined as eigenstates of the SAO. Unlike the canonical coherent states, these are not always MUS. (2) $\ket{n}$ refers to the nth eigenstate of the Hamiltonian of the bosonic (standard) harmonic oscillator, while a supercoherent state is denoted by $\ket{Z}$ with eigenvalue $z_0$. Any other symbol within a ket ($\ket{x}$) refers to the canonical coherent state with eigenvalue $x$. In particular, the symbol $\ket{\alpha}$ is used to refer to a canonical coherent state with an arbitrary eigenvalue. (3) We take $\hbar$ and $m$ to be unity; position and momentum are defined with dimensionless operators $\hat{\xi}$ and $\hat{\mu}$ defined below. (4) When operating on a superstate, any non-supersymmetric operator, e.g. $\hat{\xi}$, is externally multiplied by the $2\times 2$ identity matrix. (5) For simplicity, we use unnormalized $\ket{Z}$, although when computing physical quantities, $\braket{Z}{Z}^{-1} \ket{Z}$ is necessary. 

\section{Energy Eigenstates}

The Hamiltonian for the supersymmetric harmonic oscillator is derived in Ref. \cite{Aragone}:
\begin{equation}
  \hat{H} =
  \hat{H}_b - \frac{1}{2} \omega \sigma_3
  =
  \begin{pmatrix}
    \hat{H}_b - \frac{1}{2} \omega & 0 \\ 0 & \hat{H}_b + \frac{1}{2}\omega
  \end{pmatrix}
  =
  \begin{pmatrix}
    \omega a^+ a & 0 \\ 0 & \omega a a^+
  \end{pmatrix}
  \label{eq:Hamiltonian_Definition}
\end{equation}
where $\sigma_3$ is the diagonal Pauli matrix and $\hat{H}_b$ is the Hamiltonian defined for the bosonic (nonsupersymmetric) quantum harmonic oscillator. 
This Hamiltonian's eigenstates are given by:
\begin{equation}
  \hat{H}
  \begin{pmatrix}
    \ket{n} \\ 0
  \end{pmatrix}
  = n \omega 
  \begin{pmatrix}
    \ket{n} \\ 0
  \end{pmatrix}
\qquad
  \hat{H}
  \begin{pmatrix}
    0 \\ \ket{n-1}
  \end{pmatrix}
  = n \omega 
  \begin{pmatrix}
    0 \\ \ket{n-1}
  \end{pmatrix}
  \label{eq:Hamiltonian_Eigenstates}
\end{equation}
where $\ket{n}$ is the nth eigenstate of the Hamiltonian $\hat{H}_b \ket{n} = \left( n + \frac{1}{2} \right) \omega \ket{n}$. Yet, unlike the standard harmonic oscillator, the supersymmetric case has vanishing zero-point energy. The time-evolution of each state goes as $\exp(-i E_n t)$.

\section{Generalized Supersymmetric Annihilation Operator}

The generalized operator considered here consists only of constants and the boson annihilation operator $a$, for which $[a,a^+] = 1$. This implies that our SAO is not the most general form possible; see Ref. \cite{Lee}. Nonetheless, it is sufficiently general that we discover new supercoherent states and MUS. We construct a generalized SAO, called $\hat{A}$, by imposing the definition that it must lower the subspace of energy $n \omega$ to one of energy $(n-1) \omega$:

\begin{equation}
  \hat{A} \ket{\Psi_n}
  =
  \begin{pmatrix}
    A_{11} & A_{12} \\ A_{21} & A_{22}
  \end{pmatrix}
  \begin{pmatrix}
    \alpha_1 \ket{n} \\ \alpha_2 \ket{n-1}
  \end{pmatrix}
  =
  \begin{pmatrix}
    \alpha_1' \ket{n-1} \\ \alpha_2' \ket{n-2}
  \end{pmatrix}
  \label{eq:Superannihilation_Definition}
\end{equation}

Given that $a \ket{n} = \sqrt{n} \ket{n-1}$, this equation indicates that $\hat{A}$ has the most general form:

\begin{equation}
  \hat{A} = 
  \begin{pmatrix}
    k_1 a & k_2 \\ k_3 a^2 & k_4 a
  \end{pmatrix}
  \label{eq:Superannihilation_Form}
\end{equation}
where $k_i$ are arbitrary complex numbers. For convenience in calculations, we introduce the matrix $K$:
\begin{equation}
  K = 
  \begin{pmatrix}
    k_1 & k_2 \\ k_3 & k_4 
  \end{pmatrix}
  \label{eq:K_Definition}
\end{equation}

\section{Supercoherent States}

The supercoherent states are defined as the SAO's eigenstates:
\begin{equation}
  \hat{A} \ket{Z} = z_0 \ket{Z}
  \label{eq:Supercoherent_Definition_1}
\end{equation}
It is convenient to solve this by expanding it in the basis of energy eigenstates:
\begin{align}
  \ket{Z} = 
  \begin{pmatrix}
    \sum_{n=0}^\infty a_n \ket{n}
    \\
    \sum_{n=1}^\infty c_n \ket{n-1}
  \end{pmatrix}
  & \nonumber
  \\
  \begin{pmatrix}
    k_1 a & k_2 \\ k_3 a^2 & k_4 a
  \end{pmatrix}
  \begin{pmatrix}
    \sum_{n=0}^\infty a_n \ket{n}
    \\
    \sum_{n=1}^\infty c_n \ket{n-1}
  \end{pmatrix}
  =
  z_0
  & \begin{pmatrix}
    \sum_{n=0}^\infty a_n \ket{n}
    \\
    \sum_{n=1}^\infty c_n \ket{n-1}
  \end{pmatrix}
  \label{eq:Supercoherent_Definition_2}
\end{align}

The complex scalar $z_0$ absorbs scalar multiples of $\hat{A}$. Therefore the SAO has only three independent $k_i$, because any scalar multiple of $\hat{A}$ is indistinguishable from a multiple of $z_0$. Nonetheless, all four constants are retained in the bulk of this paper, to avoid limits of $k_i$ at $0$ and $\infty$.

The time-dependence of the system simplifies, using the time evolution $\ket{n} \to \ket{n} \exp(-i E_n t)$:
\begin{equation}
  \ket{Z(t)} =
  \begin{pmatrix}
    \sum_{n=0}^\infty a_n \ket{n} e^{-i n \omega t}
    \\
    \sum_{n=1}^\infty c_n \ket{n-1} e^{-i n \omega t}
  \end{pmatrix}
  \label{eq:Supercoherent_Time}
\end{equation}

\subsection{Recursion Relations}

Inserting the time-dependent supercoherent state into the eigensystem equation (\ref{eq:Supercoherent_Definition_2}), the summations transform into a set of recursion relations. 
\begin{align}
  \begin{cases}  
    k_1  \sqrt{n+1} a_{n+1}  + k_2  c_{n+1} = z_0 a_n \\
    k_3 \sqrt{n} \sqrt{n+1}  a_{n+1} + k_4  \sqrt{n}  c_{n+1} = z_0 c_n
  \end{cases}
  \label{eq:Recursion_Relations}
\end{align}
It is interesting that the time-dependence conveniently vanishes here, which implies that the supercoherent states have no time-dependent deformation.

We solve these recursion relations through the change of variables $ \tilde{a}_n \equiv \sqrt{n!} z_0^{1-n} a_n$ and $\tilde{c}_n \equiv \sqrt{(n-1)!} z_0^{1-n} c_n$ for $n \ge 1$. (For $a_1$, the relationships of $n=0$ show $k_1 a_1 = z_0 a_0 - k_2 c_1$. Both $a_0$ and $c_1$ are free parameters.) This reduces Eq. (\ref{eq:Recursion_Relations}) to:
\begin{equation}
  \begin{pmatrix}
    k_1 & k_2 \\ k_3 & k_4
  \end{pmatrix}
  \begin{pmatrix}
    \tilde{a}_{n+1} \\ \tilde{c}_{n+1} 
  \end{pmatrix}
  =
  \begin{pmatrix}
    \tilde{a}_n \\ \tilde{c}_n
  \end{pmatrix}
  \label{eq:Recursion_Simple}
\end{equation}

For $K$ with nonzero eigenvalues $\chi_+$ and $\chi_-$, the solution is:
\begin{align}
  \chi_\pm 
  = \frac{k_1 + k_4}{2} \pm \sqrt{ \left( \frac{k_1 - k_4}{2} \right)^2 + k_2 k_3 } 
  & = \frac{\tr(K)}{2} \pm \sqrt{ \left( \frac{\tr(K)}{2} \right)^2 - \det(K) }
  \nonumber \\ 
  S = 
  \begin{pmatrix}
    \chi_+ - k_4 & \chi_- - k_4 \\ k_3 & k_3 
  \end{pmatrix}
  \qquad & \qquad
  \beta_\pm \equiv \frac{z_0}{\chi_\pm}
  \nonumber \\
   \begin{pmatrix}
    \tilde{a}_{n+1} \\ \tilde{c}_{n+1}
  \end{pmatrix}
  = \left( K^{-1} \right)^{n} 
  \begin{pmatrix}
    \tilde{a}_1 \\ \tilde{c}_1 
  \end{pmatrix}
  & = z_0^{-n} S 
  \begin{pmatrix}
    \beta_+^n & 0 \\ 0 & \beta_-^n
  \end{pmatrix}
  S^{-1}  
  \begin{pmatrix}
    \tilde{a}_1 \\ \tilde{c}_1 
  \end{pmatrix}
  \label{eq:Diagonalization}
\end{align}

Defining $D$ as follows,
\begin{align}
  D &=
  \begin{pmatrix}
    d_{11} & d_{12} \\ d_{21} & d_{22}
  \end{pmatrix}
  = S
  \begin{pmatrix}
    \beta_+^n & 0 \\ 0 & \beta_-^n
  \end{pmatrix}
  S^{-1}
  \nonumber \\
  &= \frac{1}{\chi_+-\chi_-}
  \begin{pmatrix}
    (\chi_+ - k_4)\beta_+^n - (\chi_- - k_4)\beta_-^n
    & k_2 (\beta_+^n - \beta_-^n)
    \\ k_3 (\beta_-^n - \beta_+^n)
    & (\chi_+ - k_4)\beta_-^n - (\chi_- - k_4)\beta_+^n
  \end{pmatrix}
  \label{eq:Recursion_D}
\end{align}
the original $a_n$ and $c_n$ are found:
\begin{equation}
  \begin{pmatrix}
    a_{n+1} \\ c_{n+1}
  \end{pmatrix}
  = 
  a_0
  \begin{pmatrix}
    \frac{z_0}{k_1 \sqrt{(n+1)!}} d_{11}
    \\ \frac{z_0}{k_1 \sqrt{n!} } d_{21}
  \end{pmatrix}
  + c_1
  \begin{pmatrix}
    \frac{1}{\sqrt{(n+1)!}} \left( d_{12} - d_{11} \frac{k_2}{k_1} \right)
    \\
    \frac{1}{\sqrt{n!}} \left( d_{22} - d_{21} \frac{k_2}{k_1} \right) 
  \end{pmatrix}
  \label{eq:Recursion_Solution}
\end{equation}

This can be inserted into the eigenstate expansion in Eq. (\ref{eq:Supercoherent_Time}). The summations can be rewritten as canonical coherent states, for which a coherent state with arbitrary eigenvalue $\alpha$ is given (unnormalized, and up to a phase shift) by $\ket{\alpha} = \sum_{n=0}^\infty (n!)^{-1/2} (\alpha e^{-i \omega t})^n \ket{n}$. With this and absorbing some constants into $a_0$ and $c_1$, the supercoherent states can be expressed as any superposition of two superstates, $\ket{Z_A}$ and $\ket{Z_C}$, where $\ket{\beta_\pm}$ refer to the canonical coherent state with eigenvalue $\beta_\pm$.
\begin{align}
  \ket{Z} &= a_0 \ket{Z_A} + c_1 \ket{Z_C}
  \qquad
  \ket{Z_A} = \frac{1}{\chi_+ - \chi_-} G_A B
  \qquad
  \ket{Z_C} = \frac{1}{\chi_+ - \chi_-} G_C B
  \nonumber \\
  B &= 
  \begin{pmatrix}
    \ket{ \beta_+ e^{-i \omega t} }
    \\ \ket{ \beta_- e^{-i \omega t} }
  \end{pmatrix}
  \qquad
  G_A = 
  \begin{pmatrix}
    \chi_+ (\chi_+ - k_4)
    & - \chi_- (\chi_- - k_4)
    \\ k_3 z_0 e^{-i \omega t}
    & - k_3 z_0 e^{-i \omega t}
  \end{pmatrix}
  \nonumber \\
  G_C &=
  \begin{pmatrix}
    \chi_+ \chi_- k_2
    & - \chi_- \chi_+ k_2
    \\ z_0 e^{-i \omega t} \left[ \chi_+ k_1 - \left( k_1^2 + k_2 k_3 \right) \right] 
    & - z_0 e^{-i \omega t} \left[ \chi_- k_1 - \left( k_1^2 + k_2 k_3 \right) \right] 
  \end{pmatrix}
  \label{eq:Supercoherent_ZAC}
\end{align}

The value $z_0$ appears only in conjunction with $\exp(-i \omega t)$, so henceforth variables $z$ and $\beta_\pm$ are defined as $z = z_0 \exp(-i \omega t)$ and $\beta_\pm = z \chi_\pm^{-1}$. This rotation through the complex plane is familiar from the canonical coherent state. But this equation lacks the time-dependent constant phase shift $\exp(-i \omega t/2)$ of the coherent state, due to the vanishing energy of the ground state. 

\section{Three Families}

The supercoherent states' behavior depends on the eigenvalues $\chi_\pm$, so can be classified within the parameter space $k_i$. Specifically, the space separates into three regions: degenerate ($\chi_+ = \chi_-$), singular ($\chi_+ \chi_- = 0$, i.e. $K$ is singular), and generic (everywhere else).

In the following sections, a ``canonical supercoherent state'' refers to a superstate where each component is a scalar multiple of a particular canonical coherent state. Canonical supercoherent states are \textit{always} MUS: Consider the observables for (dimensionless) position $\hat{\xi} = (a^+ + a)/\sqrt{2}$ and momentum $\hat{\mu} = i (a^+ - a) / \sqrt{2}$. Then:
\begin{align}
  \ket{Z} = 
  \begin{pmatrix}
    c_1 \ket{\alpha} \\ c_2 \ket{\alpha}
  \end{pmatrix}
  & \qquad
  \braket{Z}{Z} =
  \left( \left| c_1\right|^2 + \left|c_2\right|^2 \right) \braket{\alpha}{\alpha}
  \nonumber \\
  \lrangle{\xi} = \sqrt{2} \Re(\alpha)
  \qquad
  \lrangle{\xi^2} = 2 \Re(\alpha)^2 + \frac{1}{2}
  & \qquad
  \lrangle{\mu} = \sqrt{2} \Im(\alpha)
  \qquad
  \lrangle{\mu^2} = 2 \Im(\alpha)^2 + \frac{1}{2}
  \label{eq:CSCS_States}
\end{align}

\subsection{Degenerate}

The necessary and sufficient condition for $\chi_+ = \chi_-$ is that $(k_1 - k_4)^2 + 4 k_2 k_3 = 0$. To solve the indeterminacy of $\ket{Z}$ in Equation (\ref{eq:Supercoherent_ZAC}), consider the case of $\chi_+ = \chi_- + \varepsilon$, for which $\lim_{\varepsilon \to 0} \chi_+ = \lim_{\varepsilon \to 0} \chi_-  \equiv \chi$. Let the function $g_{ij}(x)$ be defined for $i = A, C$ and $j = 1, 2$, such that the $(j,1)$ entry of the matrix $G_i$ is $g_{ij}(\chi_+)$ and the $(j,2)$ entry is $-g_{ij}(\chi_-)$. Hence:
\begin{align}
  & g_{A1}(x) = x^2 - k_4 x
  & \quad
  & g_{A1}'(x) = 2x - k_4
  \nonumber \\
  & g_{A2}(x) = k_3 z
  & \quad
  & g_{A2}'(x) = 0
  \nonumber \\
  & g_{C1}(x) = \chi_+ \chi_- k_2
  & \quad
  & g_{C1}'(x) = 0
  \nonumber \\
  & g_{C2}(x) = z \left[ x k_1 - \left( k_1^2 + k_2 k_3 \right) \right]
  & \quad
  & g_{C2}'(x) = z k_1
  \label{eq:Degenerate_Entries}
\end{align}

Thus the jth component of $\ket{Z_i}$ becomes:
\begin{align}
  Z_{ij} & = \sum_{n=0}^\infty \frac{\ket{n}}{\sqrt{n!}} z^n \lim_{\varepsilon \to 0} \frac{g_{ij} (\chi_- + \varepsilon) (\chi_- + \varepsilon)^{-n} - g_{ij} (\chi_-) \chi_-^{-n}}{\varepsilon}
  \nonumber \\
  &= \sum_{n=0}^\infty \frac{\ket{n}}{\sqrt{n!}} z^n \frac{d}{dx} \left(g_{ij}(x) x^{-n} \right)_{x = \chi}
  \nonumber \\
  &= \sum_{n=0}^\infty \frac{\ket{n}}{\sqrt{n!}} (z \chi^{-1})^n g_{ij}'(\chi) - \sum_{n=0}^\infty \frac{\ket{n}}{\sqrt{n!}} (z \chi^{-1})^{n-1} g_{ij}(\chi) z \chi^{-2}
  \nonumber \\
  &= g_{ij}'(\chi) \ket{\beta} - g_{ij} \chi^{-1} \beta \ket{\beta'}
  \label{eq:Degenerate_Limit}
\end{align}
where the relationship between any coherent state $\ket{\alpha}$ and its derivative $\ket{\alpha'}$ is given as in Ref. \cite{Aragone} and obeys the following relations:
\begin{equation}
  \ket{\alpha'} 
  = \frac{d}{d\alpha } \ket{\alpha}
  = a^+ \ket{\alpha}
  = \sum_{n=0}^\infty \frac{\ket{n}}{\sqrt{n!}} n \alpha^{n-1}
  \label{eq:Derivative_Coherent_State}
\end{equation}

Thus the supercoherent states for the degenerate states are linear combinations of $\ket{Z_A^d}$ and $\ket{Z_C^d}$:
\begin{align}
  & \ket{Z} = a_0 \ket{Z_A^d} + c_1 \ket{Z_C^d}
  \qquad
  \ket{Z_A^d} = G_A^d B^d
  \qquad 
  \ket{Z_C^d} = G_C^d B^d
  \nonumber \\
  & B^d = 
  \begin{pmatrix}
    \ket{\beta} \\ \ket{\beta'}
  \end{pmatrix}
  \qquad
  G_A^d = 
  \begin{pmatrix}
    \chi_1
    & -(\chi - k_4)\beta
    \\ 0 
    & -k_3 \beta^2
  \end{pmatrix}
  \qquad
  G_C^d =
  \begin{pmatrix}
    0
    & -k_2 \chi \beta
    \\ k_1 \chi \beta 
    & -\left( \frac{k_4^2 - k_1^2}{4} \right) \beta^2
  \end{pmatrix}
  \label{eq:Degenerate_States}
\end{align}

A linear combination of these states presents one canonical supercoherent state in the subspace:
\begin{equation}
  \ket{Z_{MUS}^d} =
  \begin{pmatrix}
    - k_1 k_2  \chi \ket{\beta}
    \\ k_1 \frac{k_1^2 - k_4^2 }{4} \beta \ket{\beta}
  \end{pmatrix}
  \label{eq:Degenerate_ZS}
\end{equation}

\subsection{Singular}

If $K$ is singular, then a zero eigenvalue prevents matrix decomposition. The necessary and sufficient condition for this is $k_1 k_4 = k_2 k_3$. Then the recursion relations become:
\begin{equation}
  \begin{pmatrix}
    k_1 & k_2 \\ k_1 k_4 / k_2 & k_4
  \end{pmatrix}
  \begin{pmatrix}
    \tilde{a}_{n+1} \\ \tilde{c}_{n+1}
  \end{pmatrix}
  = 
  \begin{pmatrix}
    \tilde{a}_n \\ \tilde{c}_n
  \end{pmatrix}
  \label{eq:Recursion_Singular}
\end{equation}
which are solved by a one-dimensional supercoherent space; after eliminating spurious constants, this becomes:
\begin{equation}
  c_{n+1} = \frac{k_4 \beta}{k_2} a_n = \frac{k_4 \beta a_0 }{k_2 \sqrt{n!} } \beta^{n}
  \qquad
  \beta \equiv \frac{z}{k_1 + k_4}
  \qquad
  \ket{Z_s} = 
  \begin{pmatrix}
    k_2 \ket{\beta}
    \\ k_4 \beta \ket{\beta}
  \end{pmatrix}
  =
  \begin{pmatrix}
    k_1 \ket{\beta} \\ k_3 \beta \ket{\beta}
  \end{pmatrix}
  \label{eq:Singular_State}
\end{equation}
where the last equality assumes scalar multiplication by the constant $k_1/k_2$. This set of states is unique in two ways: (1) This is a one-dimensional supercoherent space, rather than the two-dimensional space found in the other regions of the $k_i$ parameter space. (2) Therefore, not only does a canonical supercoherent state exist, but all supercoherent states in this region are canonical supercoherent states. Due to this, all supercoherent states in this region are MUS.

\subsection{Generic}

When $K$ has nonzero nonequal eigenvalues, the above basis $\left\{ \ket{Z_A}, \ket{Z_C} \right\} $ is easily calculable. In this space there are two canonical supercoherent states, which we choose as the new basis vectors for the supercoherent space:
\begin{equation}
  \ket{Z_\pm} =
  \begin{pmatrix}
    k_2 \chi_\pm \ket{\beta_\pm}
    \\
    \left(\chi_\pm - k_1 \right) z \ket{\beta_\pm}
  \end{pmatrix}
  \label{eq:Generic_States}
\end{equation}

The parameter space of $\left\{k_1, k_2, k_3, k_4\right\}$ can be changed to a parameter space of $\left\{\chi_+ , \chi_-, k_1, k_2 \right\}$. The role of $k_2$ consists of controlling the ratio between the two components of $\ket{Z}$.

The basis vectors, as canonical supercoherent states, always minimize uncertainty. In fact, there are no canonical supercoherent states in the subspace other than the basis states themselves. Any such state $\ket{\beta_*}$ would have to fulfill the equation $\ket{\beta_*} = A_+ \ket{\beta_+} + A_- \ket{\beta_-}$ for $\beta_* \ne \beta_\pm$. By expanding the canonical coherent states in the eigenstate basis, the necessary and sufficient condition appears as:
\begin{equation}
  A_+ \left( \left| \frac{\beta_+}{\beta_*} \right| e^{i \theta_+} \right)^n
  +
  A_- \left( \left| \frac{\beta_-}{\beta_*} \right| e^{i \theta_-} \right)^n
  = 1 
  \quad \forall n \in \mathbb{N}
  \label{eq:CSCS_Superposition}
\end{equation}
 
To satisfy the limit of $n \to \infty$ while $\beta_* \ne \beta_\pm$, this requires $ | \beta_+ | = | \beta_- |$. Then each term of the left-hand-side has the same magnitude, so a sum of unity for all $n$ requires that $\theta_+ = \theta_-$, which cannot apply to the non-degenerate case $\chi_+ \ne \chi_-$.

\section{Superpositions and Uncertainties}

The basis states of the generic supercoherent subspace are canonical supercoherent states, so are always MUS. However, not all states of that subspace minimize uncertainty. This section describes interesting behavior for a particular one-parameter family of supercoherent states. We describe the calculations for these uncertainties, then analyze the theoretical conditions that lead to bounded uncertainty.

\subsection{A Numerical Example}

To gain insight into the behavior of the non-intuitive 3-parameter supercoherent states, we build a 1-parameter example for exploration. We consider the following annihilation matrix, that for $\theta=0$ reduces to the original SAO of Ref. \cite{Aragone}:
\begin{equation}
  \hat{A}_\theta = 
  \begin{pmatrix}
    a & \cos \theta \\ a^2 \sin \theta & a
  \end{pmatrix}
  \label{eq:Superannihilation_Theta}
\end{equation}

The supercoherent states separate into multiple regions:
\begin{enumerate}
  \item In the region $0 < \theta < \pi/2$, the eigenvalues are real and distinct. As proven below, this implies that the uncertainty is bounded for all values of $z$.

  \item In the region $\pi/2 < \theta < \pi$, the eigenvalues are imaginary and have the same norm. As proven below, this leads to uncertainty diverging with some power of $z$.

  \item Degenerate states occur at $ \theta = n \pi / 2$. These are analyzed above.

  \item The uncertainty cycles over a period of $\pi$, so we henceforth ignore $\theta > \pi$.
\end{enumerate}

Now consider the state mixing the two basis states as follows:
\begin{equation}
  \ket{Z_\theta} = \frac{1}{\sqrt{2}} \left( \ket{Z_+} + \ket{Z_-} e^{i \pi/4} \right)
  \label{eq:Z_theta_mixed}
\end{equation}

\begin{figure}[htb]
  \begin{center}
    \subfloat{
      \includegraphics[width=0.4\textwidth]{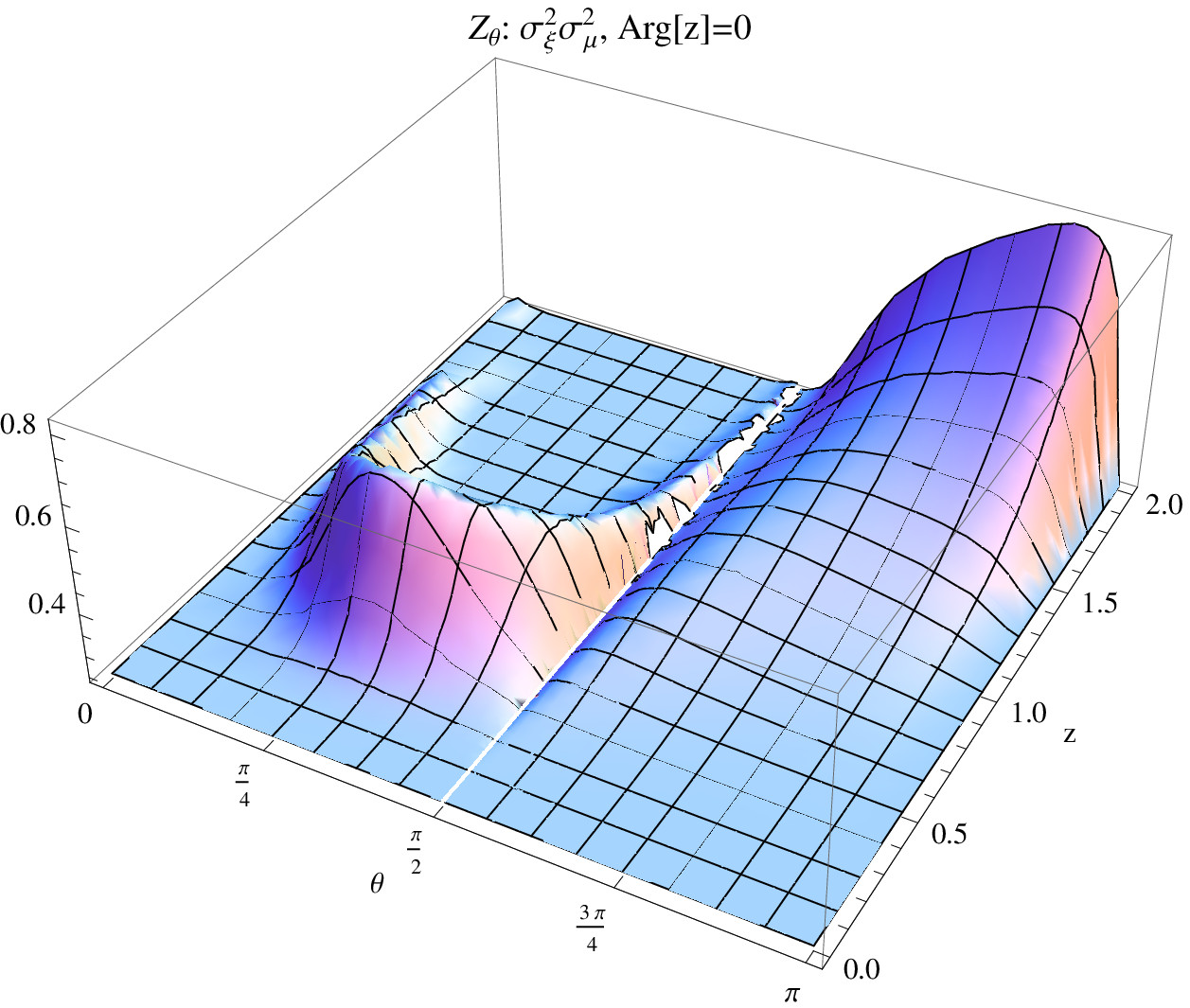}
    }
    \subfloat{
      \includegraphics[width=0.4\textwidth]{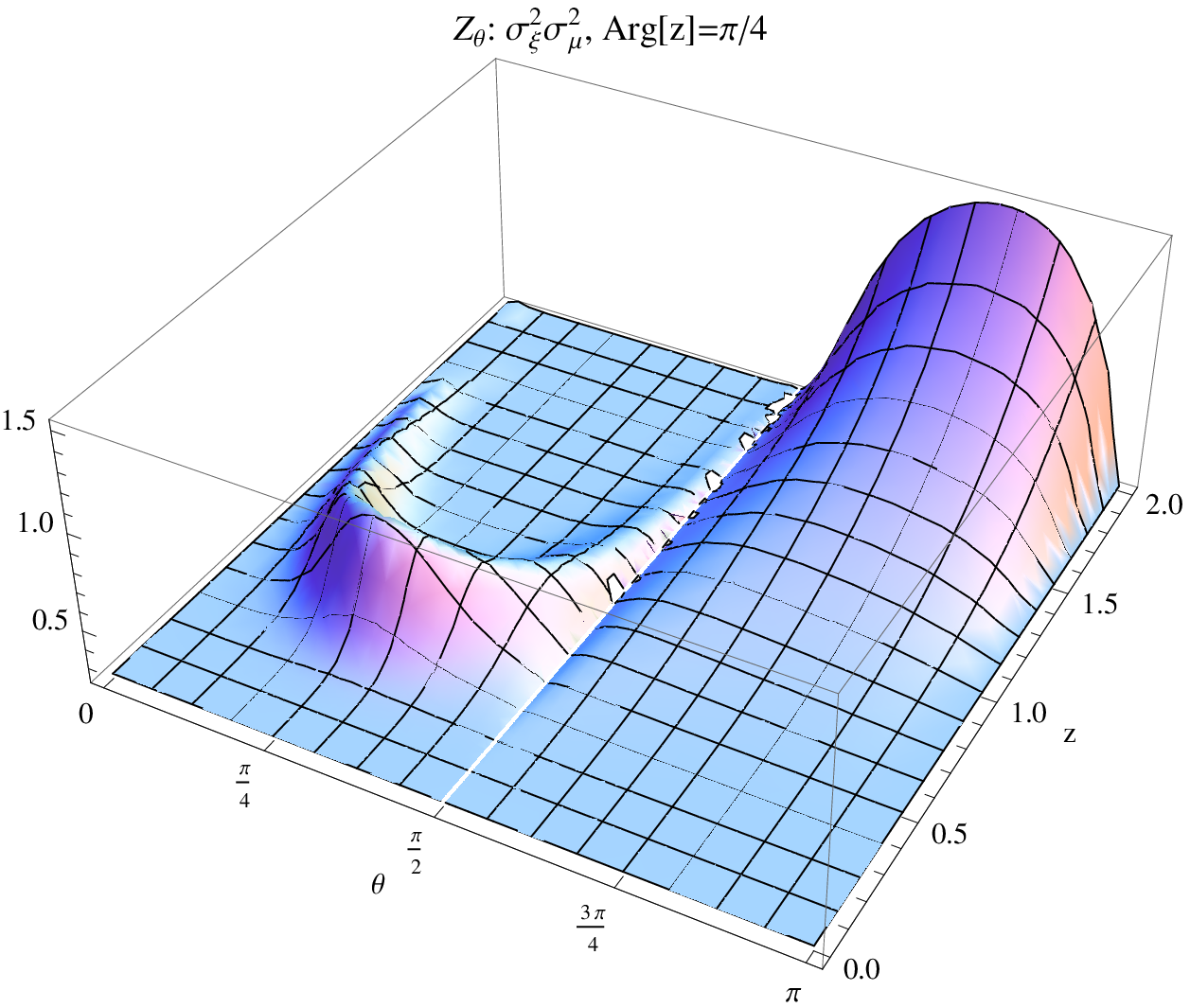}
    }
  \end{center}
  \caption{Uncertainty of the supercoherent state $\ket{Z_\theta}$, which mixes the two basis states ($\ket{Z_\theta} = 2^{-1/2} \ket{Z_+} + 2^{-1/2} e^{i \pi/4} \ket{Z_-}$). The uncertainty separates into two regions: (1) In $0 < \theta < \pi/2$, the uncertainty is bounded. The maximum is $0.83$, at approximately $z = 0.5 e^{i \pi/4}, \theta = \pi/4$. (2) In $ \pi/2 < \theta < \pi$, the uncertainty diverges with $z$. The rate of divergence depends upon the phase of $z$: We show that for $\arg(z) = 0$, it diverges as $z^2$, while for $\arg(z) = \pi/4$, it diverges as $z^4$. }
  \label{fig:Ztheta}
\end{figure}

Using the calculations described below, the uncertainty is calculated numerically for the non-degenerate regions in Figure \ref{fig:Ztheta}. The first two regions have vastly different behavior: In the region of real eigenvalues ($0 < \theta < \pi/2$) the uncertainty reaches a maximum of approximately $0.83$ (at $z \sim 0.5 e^{i \pi/4}, \theta \sim \pi/4$, compared to a minimum $\sigma_\xi^2 \sigma_\mu^2 = 1/4$), then returns to saturated uncertainty as $z$ increases. In the other region, $\pi/2 < \theta < \pi$, the uncertainty diverges as a power of $z$. The figures show that the rate of divergence depends on the phase of $z$. As shown below, if $m$ is an integer, the uncertainty of $\arg(z) = m \pi/2$ diverges as $z^2$, while for $\arg(z) \ne m \pi/2 $ it diverges as $z^4$.

\subsection{Generic}

The calculations of uncertainty are computationally straightforward, if a bit lengthy. Consider a mixture of the aforementioned basis states $\ket{Z_\pm}$, with the parameters $\eta$ and $\lambda$ determining the extent of mixing:
\begin{align}
  \ket{Z_m} =
  \begin{pmatrix}
    \gamma_{1+} \ket{\beta_+} + \gamma_{1-} \ket{\beta_-}
    \\ \gamma_{2+} \ket{\beta_+} + \gamma_{2-} \ket{\beta_-}
  \end{pmatrix}
  & \qquad
  \gamma_{1+} = k_2 \chi_+ \cos\eta
  \qquad
  \gamma_{1-} = k_2 \chi_- e^{i\lambda} \sin\eta
  \nonumber \\ 
  & \qquad
  \gamma_{2+} = ( \chi_+ - k_1)  \cos\eta
  \qquad
  \gamma_{2-} = ( \chi_- - k_1)  e^{i\lambda} \sin\eta
  \nonumber \\ 
  \Gamma_+ = \left|\gamma_{1+}\right|^2 + \left|\gamma_{2+} z \right|^2 
  \qquad
  & \Gamma_- = \left|\gamma_{1-}\right|^2 + \left|\gamma_{2-} z \right|^2 
  \qquad
  \Gamma_{+-} = \gamma_{1+}^* \gamma_{1-} + \gamma_{2+}^* \gamma_{2-} |z|^2
  \label{eq:Mixed_State}
\end{align}

For any supersymmetric diagonal observable $\hat{c}$, the expectation value is:
\begin{equation}
  \lrangle{c} =
  \frac{ \braket{Z_m}{\hat{c} Z_m} }{ \braket{Z_m}{Z_m} }
  \qquad
  \braket{Z_m}{\hat{c} Z_m} = \Gamma_+ \braket{\beta_+}{\hat{c}\beta_+} + \Gamma_- \braket{\beta_-}{\hat{c}\beta_-} + 2 \Re \left( \Gamma_{+-} \braket{\beta_+}{\hat{c} \beta_-} \right)
  \label{eq:Mixed_Observable}
\end{equation}

To calculate these brakets, consider the following results, for the observables $\hat{\xi}$ and $\hat{\mu}$ defined above. For any two canonical coherent states $\ket{\alpha_1}$ and $\ket{\alpha_2}$,

\begin{align}
  \braket{\alpha_1}{\alpha_2}
  &= \sum_{n=0}^\infty 
    \left( \frac{\overline{\alpha_1^n}}{\sqrt{n!}} \bra{n} \right)
    \left( \frac{  \alpha_2^n}{\sqrt{n!}} \ket{n} \right)  
  = \exp \left( \overline{\alpha_1}  \alpha_2 \right)
  \equiv \epsilon
  \nonumber \\
  \braket{\alpha_1}{\xi \alpha_2} 
  &= \frac{1}{\sqrt{2}} \left(\alpha_2 + \overline{\alpha_1} \right) \epsilon
  \nonumber \\
  \braket{\alpha_1}{\xi ^2 \alpha_2}
  &= \frac{1}{2} \left[ \left( \alpha_2 + \overline{\alpha_1} \right)^2 + 1 \right] \epsilon
  \nonumber \\
  \braket{\alpha_1}{\mu \alpha_2} 
  &= \frac{-i}{\sqrt{2}} \left( \alpha_2 - \overline{\alpha_1} \right) \epsilon
  \nonumber \\
  \braket{\alpha_1}{\mu ^2 \alpha_2}
  &= \frac{1}{2} \left[ - \left( \alpha_2 - \overline{\alpha_1} \right)^2 + 1 \right] \epsilon
\label{eq:Mixed_Brakets}
\end{align}

These relationships between $\xi$ and $\mu$ indicate that $\lrangle{\mu}$ and $\lrangle{\mu^2}$ are equivalent to $\lrangle{\xi}$ and $\lrangle{\xi^2}$ under the transformation $\alpha_j \to -i \alpha_j$. Given that the physically relevant values of $\alpha$ are $\beta_\pm = z / \chi_\pm$, it is apparent that the momentum expectation values can be calculated under the transformation $z \to -i z$. As with the canonical coherent state, the rotation of the eigenvalue $z$ through the complex plane is the mathematical representation of the state's rotation through phase space.

Therefore, the explicit expressions for variance in position and momentum become:
\begin{align}
  \sigma_\xi^2 &= 
  \frac{
    \Gamma_{+} \frac{1}{2}[(\beta_+ + \beta_+^*)^2+1] e^{|\beta_+|^2} +
    \Gamma_{-} \frac{1}{2}[(\beta_- + \beta_-^*)^2+1] e^{|\beta_-|^2} + 
    2 \Re \left[ \Gamma_{+-} \frac{1}{2}[(\beta_- + \beta_+^*)^2+1] e^{\beta_+^* \beta_-} \right]
  }{
    \Gamma_{+} e^{|\beta_+|^2} +
    \Gamma_{-} e^{|\beta_-|^2} + 
    2 \Re \left[ \Gamma_{+-} e^{\beta_+^* \beta_-} \right]
  }
  \nonumber \\ \qquad &
  -
  \left( \frac{
    \Gamma_{+} \frac{1}{\sqrt{2}}(\beta_+ + \beta_+^*)  e^{|\beta_+|^2} +
    \Gamma_{-} \frac{1}{\sqrt{2}}(\beta_- + \beta_-^*) e^{|\beta_-|^2} + 
    2 \Re \left[ \Gamma_{+-} \frac{1}{\sqrt{2}}(\beta_- + \beta_+^*) e^{\beta_+^* \beta_-} \right]
  }{
    \Gamma_{+} e^{|\beta_+|^2} +
    \Gamma_{-} e^{|\beta_-|^2} + 
    2 \Re \left[ \Gamma_{+-} e^{\beta_+^* \beta_-} \right]
  } \right) ^2
  \nonumber \\
  \sigma_\mu^2 &= 
  \frac{
    \Gamma_{+} \frac{1}{2}[- (\beta_+ - \beta_+^*)^2+1] e^{|\beta_+|^2} +
    \Gamma_{-} \frac{1}{2}[- (\beta_- - \beta_-^*)^2+1] e^{|\beta_-|^2} + 
    2 \Re \left[ \Gamma_{+-} \frac{1}{2}[- (\beta_- - \beta_+^*)^2+1] e^{\beta_+^* \beta_-} \right]
  }{
    \Gamma_{+} e^{|\beta_+|^2} +
    \Gamma_{-} e^{|\beta_-|^2} + 
    2 \Re \left[ \Gamma_{+-} e^{\beta_+^* \beta_-} \right]
  }
  \nonumber \\ \qquad &
  -
  \left( \frac{
    \Gamma_{+} \frac{-i}{\sqrt{2}}(\beta_+ - \beta_+^*)  e^{|\beta_+|^2} +
    \Gamma_{-} \frac{-i}{\sqrt{2}}(\beta_- - \beta_+^*) e^{|\beta_-|^2} + 
    2 \Re \left[ \Gamma_{+-} \frac{1}{\sqrt{2}}(\beta_- - \beta_+^*) e^{\beta_+^* \beta_-} \right]
  }{
    \Gamma_{+} e^{|\beta_+|^2} +
    \Gamma_{-} e^{|\beta_-|^2} + 
    2 \Re \left[ \Gamma_{+-} e^{\beta_+^* \beta_-} \right]
  } \right) ^2
  \label{eq:Mixed_Uncertainties}
\end{align}

\subsection{Boundedness}

As shown, some annihilation operators have bounded uncertainty for all values of $z$, e.g. $\hat{A}_\theta$ for $0 < \theta < \pi/2$, while others have unbounded uncertainty, e.g. $\hat{A}_\theta$ for $\pi / 2 < \theta < \pi$. In this section, we show the conditions necessary for each case. 

The uncertainty product $\sigma_\xi \sigma_\mu$ given by Eq. (\ref{eq:Mixed_Uncertainties}) is considered to be bounded if, for a given operator $\hat{A}$, all eigenstates have finite uncertainty. Due to finite and nonzero normalization, the only relevant condition is the boundary conditions of $z$.

For $z \to 0$, it is apparent that due to cancellations, the uncertainty reaches the Heisenberg minimum, $\sigma_\xi^2 = \sigma_\mu^2 = 1/2$. For $z \to \infty$, the uncertainty differs between two regions.

For $|\chi_+| \ne |\chi_-|$, the uncertainties are dominated by either the $|\beta_+|^2$ or the $|\beta_-|^2$ exponential. Once the other two exponentials drop away, this region reaches the minimum-uncertainty limit $\sigma_\xi^2 = \sigma_\mu^2 = 1/2$.

However, if $|\chi_+| \ne |\chi_-|$, no single term dominates; simultaneously, avoiding the degenerate case requires $\chi_+ \ne \chi_-$. These two conditions are satisfied only if $\chi_\pm$ are complex. For ease of calculation, consider the following equations: 
\begin{equation}
  \chi_\pm = \chi e^{i\phi_\pm} 
  \qquad
  z = z_0 e^{-i\omega t} 
  \qquad
  \beta_\pm = \beta_0 e^{i(-\omega t - \phi_\pm)}
  \qquad
   |\beta_\pm|^2 = \beta_0^2 
  \label{eq:Mixed_Boundedness1}
\end{equation}

The $\beta_+^* \beta_-$ exponential grows slowest, as its real part is less than those of the other exponentials. Therefore, the uncertainties reduce to:
\begin{align}
  \sigma_\xi^2 &=
  \frac{1}{2} +
  \frac{\Gamma_+ \Gamma_-}{(\Gamma_+ + \Gamma_-)^2} 8 \beta_0^2
  \sin^2 \left(\frac{\phi_+ - \phi_-}{2}\right)
  \sin^2 \left(\omega t + \frac{\phi_+ + \phi_-}{2}\right)
  \nonumber \\
  \sigma_\mu^2 &=
  \frac{1}{2} +
  \frac{\Gamma_+ \Gamma_-}{(\Gamma_+ + \Gamma_-)^2} 8 \beta_0^2
  \sin^2 \left(\frac{\phi_+ - \phi_-}{2}\right)
  \cos^2 \left(\omega t + \frac{\phi_+ + \phi_-}{2}\right)
  \label{eq:Mixed_Boundedness2}
\end{align}

Given that (by definition) $\Gamma$ and $\beta_0^2$ grow as $z_0^2$ for $z_0 \to \infty$, it is apparent that $\sigma_\xi^2$ diverges as $z_0^2$ unless $2 \omega t + \phi_+ + \phi_- = n \pi $ for even $n$; similarly, $\sigma_\mu^2$ diverges as $z_0^2$ unless $2 \omega t + \phi_+ + \phi_- = m \pi $ for odd $m$. If both $k_1$ and $k_4$ are real, then $\chi_\pm$ are complex conjugates, for which $\phi_+ + \phi_- = 0$. 

These conclusions are in agreement with Figure \ref{fig:Ztheta}, where $\sigma_\mu^2$ diverged as $z^2$ for both $\omega t = \arg(z) = 0, \pi/4$, but $\sigma_\xi^2$ diverged as $z^2$ for only $\omega t = \pi/4$ but was zero for $\omega t = 0$.

Therefore, in general, non-degenerate nonsingular supercoherent states have unbounded uncertainty $\sigma_\xi \sigma_\mu$ if and only if $|\chi_+| = |\chi_-|$. For $k_1, k_4 \in \mathbb{R}$, this is equivalent to the condition $(k_1 - k_4)^2 + 4 k_2 k_3 < 0$. Otherwise it is bounded, but it is difficult to predict the numerical limit on the uncertainty.

\section{Conclusions}

This work extended the supercoherent states derived in Ref. \cite{Aragone}, focusing on their uncertainty and searching for MUS. We generalized the supersymmetric annihilation operator to a family of operators that lower the subspace of energy $E_n$ to the subspace of energy $E_{n-1}$. The eigenvalues of these operators are the new supercoherent states, for which we detected many previously unidentified states that saturate the uncertainty principle and others of bounded uncertainty. The supercoherent states, like the canonical coherent states, suffer no time-dependent deformation.

Figure \ref{fig:parameter_space} depicts our results graphically. The supersymmetric annihilation operator depends on four parameters, $k_1, k_2, k_3, k_4$, but one constant is absorbed into the eigenvalue $z_0$. If we now consider only real entries in $K$, we can graph the new $\mathbb{R}^3$ subspace in Figure \ref{fig:parameter_space}, with a few significant regions:

\begin{figure}[htb]
  \begin{center}
    \includegraphics{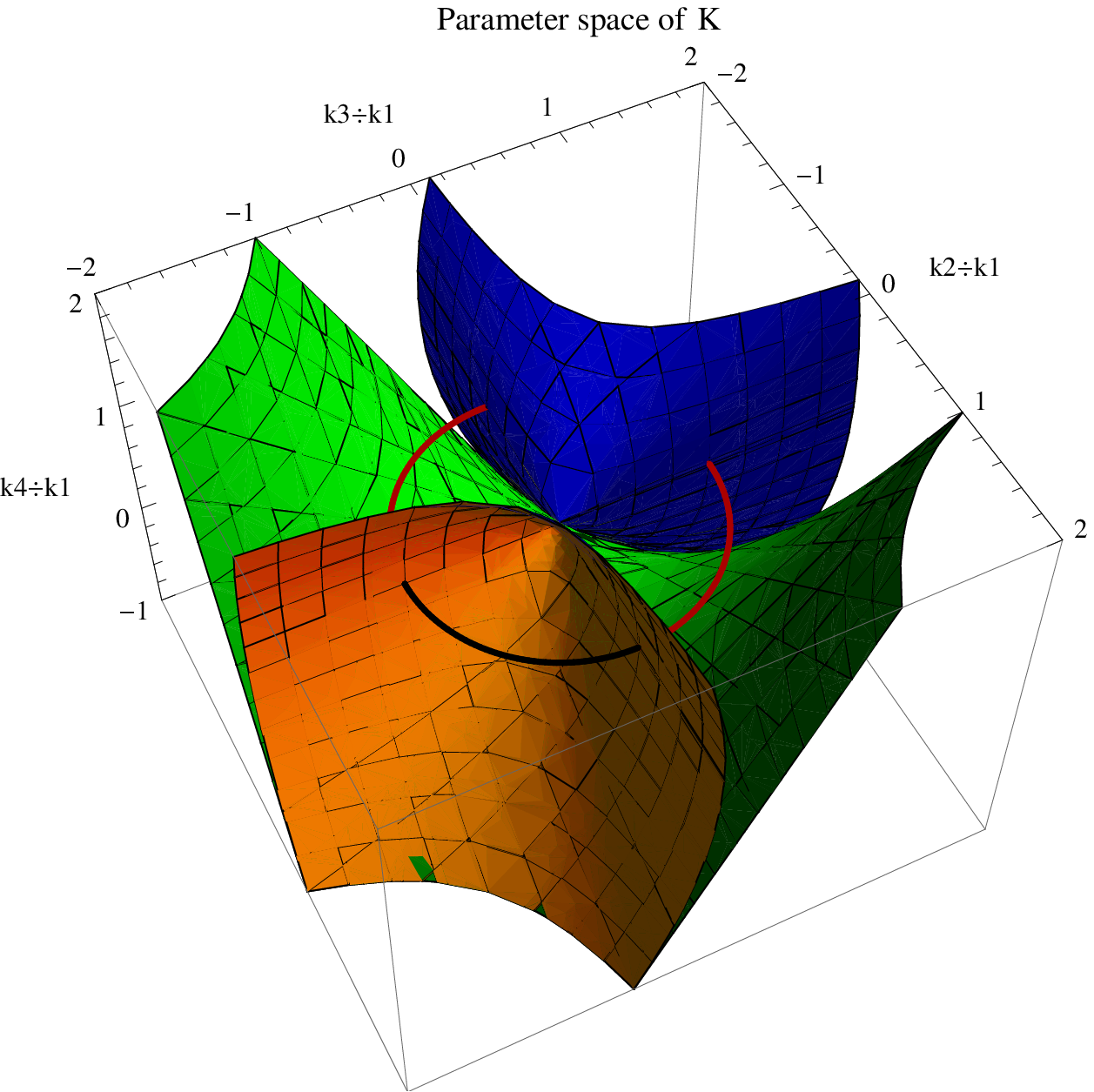}
  \end{center}
  \caption{Graph of the parameter space of the supercoherent states. The blue-yellow surface describes the degenerate case, while the green surface is the singular matrix. The red-black circle describes the family $\hat{A}_\theta$, for which the red portions have bounded uncertainty for all eigenstates while the black portions have unbounded uncertainty for most eigenstates. All parameters $k_i$ are assumed to be real and divided by $k_1$, as the supersymmetric annihilation matrix actually occupies a three-dimensional space. }
  \label{fig:parameter_space}
\end{figure}

\begin{enumerate}
  \item The blue-yellow surfaces in Figure \ref{fig:parameter_space} represent the degenerate case, where $K$ has one repeated eigenvalue $\chi$. The supercoherent subspace here is two-dimensional, with one MUS.

  \item The green surface represents the singular case, where $K$ has at least one eigenvalue equal to $0$. This supercoherent subspace is one-dimensional, which is always MUS.

  \item The solid red line is the generic family described by $\hat{A}_\theta$ for $n \pi < \theta < (2n+1) \pi/2$ for integer $n$. This has bounded uncertainty, never larger than $0.83$, as described above for $n=0$. In fact, the entire region between the degenerate surfaces has real, distinct eigenvalues, so has bounded uncertainty. The supercoherent subspace for the generic states is two-dimensional, with two basis MUS.

  \item The dashed black line is the generic family described by $\hat{A}_\theta$ for $(2n-1) \pi /2 < \theta < n \pi$. Its uncertainty is unbounded with $z$, as described above for $n=1$. The entire region outside the degenerate surfaces has $(k_1 - k_4)^2 + 4 k_2 k_3 < 0$, so (for $k_1, k_4 \in \mathbb{R}$) the eigenvalues have the same norms, leading to unbounded uncertainties. The supercoherent subspace is also two-dimensional with two basis MUS.
\end{enumerate}

\section{Acknowledgments}

We acknowledge the funding of the Kressel Research Scholarship, Gamson Fund, and Honors Program of Yeshiva University. MK thanks Yitzhak Kornbluth for fruitful discussions.


\begin{thebibliography}{99}

  \bibitem{Angelova} M. Angelova and V. Hussin ``Generalized and Gaussian coherent states for the Morse potential,'' \textit{J. Phys. A: Math. Theor.} 41 (2008) 304016

  \bibitem{Aragone} C. Aragone and F. Zypman ``Supercoherent States,'' \textit{J. Phys. A: Math. Gen.} 19 (1986) 2267-2279

  \bibitem{Berube} Y. Berube-Lauziere and V. Hussin ``Comments on the Definitions of Coherent States for the SUSY Harmonic Oscillator,'' \textit{J. Phys. A: Math. Gen.} 26 (1993) 6271-6275

  \bibitem{Cherbal} O. Cherbal, M. Drir, M. Maamache, and D.A. Trifonov ``Supersymmetric Extension of Non-Hermitian su(2) Hamiltonian and Supercoherent States,'' \textit{SIGMA} 6 (2010) 096

  \bibitem{Dodonov} V.V. Dodonov ``'Nonclassical' states in quantum optics: a 'squeezed' review of the first 75 years,'' \textit{J. Opt. B: Quantum Semiclass. Opt.} 4 (2002) R1-R33

  \bibitem{Fatyga} B.W. Fatyga, V.A. Kostelecky, M.M. Nieto, and D.R. Traux ``Supercoherent States,'' \textit{Phys. Rev. D} 43:4 (1991) 1403-1412

  \bibitem{Fernandez} D.J. Fernandez, V. Hussin, and O. Rosas-Ortiz ``Coherent States for Hamiltonians Generated by Supersymmetry,'' \textit{J. Phys. A: Math. Theor.} 40 (2007) 6491-6511

  \bibitem{Lee} Y. Lee, D.J. Kouri, and D.K. Hoffman ``Minimum uncertainty wavelets in non-relativistic super-symmetric quantum mechanics,'' \textit{Journal of Mathematical Chemistry} 49:1 (2011) 12-34

  \bibitem{Maassen} H. Maassen and J.B.M. Uffink ``Generalized Entropic Uncertainty Relations,'' \textit{Phys. Rev. Lett.} 60:12 (1988) 1103-1106

  \bibitem{Nieto} M.M. Nieto and L.M. Simmons Jr. ``Coherent states for general potentials. I. Formalism,'' \textit{Phys. Rev. D} 20:6 (1979) 1321-1331

\end{thebibliography}
\end{document}